%% file: paper.tex
\documentclass{llncs}

\usepackage{amssymb, amsmath, amsfonts, xfrac, mathabx}
\usepackage{mathtools}

\def\R{{\rm I\!R}}

\usepackage{url}
\urldef{\mailsa}\path|christoph.jud@unibas.ch|
\usepackage{siunitx}
\DeclareSIUnit\pixel{px}

\usepackage{tikz}
\usepackage{pgfplots}
\usetikzlibrary{patterns}
\usetikzlibrary{spy}
\pgfplotsset{compat=1.8}

\usepackage{xcolor}
\usepackage{tikz}
\usetikzlibrary{positioning,fadings,through,calc,decorations.pathmorphing,3d}
\usepackage{pgfplots}
\usepgfplotslibrary{groupplots}
\usetikzlibrary{spy}
\pgfplotsset{compat=1.8}
\usepackage{pgfplotstable}
\usepackage{filecontents}
\usetikzlibrary{positioning}
\usetikzlibrary{arrows}
\usetikzlibrary{shapes}

\usepackage{ifthen}

\pgfmathsetmacro{\cubex}{4}
\pgfmathsetmacro{\cubey}{4}
\pgfmathsetmacro{\cubez}{2}

\pgfmathsetmacro{\shiftx}{2.1}
\pgfmathsetmacro{\shifty}{6.3}

\colorlet{facea}{gray!10}
\colorlet{faceb}{gray!30}
\colorlet{facec}{gray!20}

\colorlet{base}{yellow}
\colorlet{side}{yellow!30}

\colorlet{base1}{cyan}
\colorlet{side1}{cyan!30}

\pgfmathsetmacro{\ra}{0.2}
\pgfmathsetmacro{\rb}{0.14}

\pgfmathsetmacro{\xdeg}{45}
\pgfmathsetmacro{\xx}{cos(\xdeg)}
\pgfmathsetmacro{\xy}{sin(\xdeg)}

\pgfmathsetmacro{\ydeg}{180}
\pgfmathsetmacro{\yx}{cos(\ydeg)}
\pgfmathsetmacro{\yy}{sin(\ydeg)}

\pgfmathsetmacro{\zdeg}{90}
\pgfmathsetmacro{\zx}{cos(\zdeg)}
\pgfmathsetmacro{\zy}{sin(\zdeg)}

\newcommand{\tdcylxy}[7]{
  \path (1,0,0);
  \pgfgetlastxy{\cylxx}{\cylxy}
  \path (0,1,0);
  \pgfgetlastxy{\cylyx}{\cylyy}
  \path (0,0,1);
  \pgfgetlastxy{\cylzx}{\cylzy}
  \pgfmathsetmacro{\cylt}{(\cylzy * \cylyx - \cylzx * \cylyy)/ (\cylzy * \cylxx - \cylzx * \cylxy)}
  \pgfmathsetmacro{\ang}{atan(\cylt)}
  \pgfmathsetmacro{\ct}{1/sqrt(1 + (\cylt)^2)}
  \pgfmathsetmacro{\st}{\cylt * \ct}
  \filldraw[thick, black, fill=#6] (#4*\ct+#1,#4*\st+#2,#3) -- ++(0,0,#5) arc[start angle=\ang,delta angle=180,radius=#4] -- ++(0,0,-#5) arc[start angle=\ang+180,delta angle=180,radius=#4];
  \filldraw[thick, black, fill=#7] (#1,#2,#3) circle[radius=#4];
}

\definecolor{hsv1}{rgb}{1.00, 0.00, 0.00}
\definecolor{hsv2}{rgb}{1.00, 0.75, 0.00}
\definecolor{hsv3}{rgb}{0.50, 1.00, 0.00}
\definecolor{hsv4}{rgb}{0.00, 1.00, 0.25}
\definecolor{hsv5}{rgb}{0.00, 1.00, 1.00}
\definecolor{hsv6}{rgb}{0.00, 0.25, 1.00}
\definecolor{hsv7}{rgb}{0.50, 0.00, 1.00}
\definecolor{hsv8}{rgb}{1.00, 0.00, 0.75}

\pgfdeclarelayer{background}    
\pgfsetlayers{background,main}  
\usepgfplotslibrary{
  dateplot,
  groupplots,
  statistics,
}

\pgfplotsset{
  compat=1.13,
  boxplot/draw direction=y,
}

\usepackage{pgfplotstable}
\pgfdeclarelayer{background}
\pgfsetlayers{background,main}

\usepackage{graphicx, tabularx}

\usepackage{eso-pic}

\usepackage{subfig}
\usepackage{hyperref}
\usepackage{adjustbox}

\newcommand*{\affmark}[1][*]{\textsuperscript{#1}}

\begin{document}

\mainmatter

\title{Accelerated Motion-Aware MR Imaging\\ via Motion Prediction from K-Space Center}
\titlerunning{Accelerated Motion-Aware MR Imaging }

\author{Christoph~Jud\affmark[1] \and Damien~Nguyen\affmark[1,2] \and Alina~Giger\affmark[1] \and\\ Robin~Sandk\"{u}hler\affmark[1] \and Miriam~Krieger\affmark[3] \and Tony~Lomax\affmark[3] \and Rares~Salomir\affmark[4]\and Oliver~Bieri\affmark[1,2] \and Philippe~C.~Cattin\affmark[1]}

\institute{\affmark[1]Department of Biomedical Engineering, University of Basel, Switzerland\\
\affmark[2]Department of Radiology, Division of Radiological Physics, University Hospital Basel, Switzerland\\
\affmark[3] Paul Scherrer Institute (PSI), Center for Proton Therapy, Villigen, Switzerland\\
\affmark[4] Image Guided Interventions Laboratory, University of Geneva, Geneva, Switzerland
\mailsa}

\maketitle

\begin{abstract}
Motion has been a challenge for magnetic resonance (MR) imaging ever since the MR has been invented. Especially in volumetric imaging of thoracic and abdominal organs, motion-awareness is essential for reducing motion artifacts in the final image. A recently proposed MR imaging approach copes with motion by observing the motion patterns during the acquisition. Repetitive scanning of the k-space center region enables the extraction of the patient motion while acquiring the remaining part of the k-space. Due to highly redundant measurements of the center, the required scanning time of over \SI{11}{\minute} and the reconstruction time of \SI{2}{\hour} exceed clinical applicability though. We propose an \emph{accelerated} motion-aware MR imaging method where the motion is inferred from small-sized k-space center patches and an initial training phase during which the characteristic movements are modeled. Thereby, acquisition times are reduced by a factor of almost $2$ and reconstruction times by two orders of magnitude. Moreover, we improve the existing motion-aware approach with a systematic temporal shift correction to achieve a sharper image reconstruction. We tested our method on $12$ volunteers and scanned their lungs and abdomen under free breathing. We achieved equivalent to higher reconstruction quality using the motion-prediction compared to the slower existing approach.

\end{abstract}


\section{Introduction}
Patient motion during magnetic resonance (MR) acquisitions poses challenges for the MR imaging process. While the MR acquisition space, called k-space, is measured, movements induce inconsistencies which lead to motion artifacts in the final image. This has severe clinical implications. For example, in liver MR examinations, motion artifacts are the leading cause for repeating an MR acquisition~\cite{schreiber2017frequency}. This involves considerable costs as shown in a neuro-radiological study about MR {{examinations~\cite{andre2015toward}}}. The estimated costs associated with motion artifacts amount up to \$115,000 per scanner per year. For abdominal and thoracic MR examinations the expenses are expected to be even higher.
Moreover, methods which can reduce motion artifacts are essential for the clinics not only because they help to cut costs but also because they reduce potential impacts on patient safety and the risks associated with misinterpretations of motion-degenerated MR images.

In~\cite{jud2018motion}, a motion-aware MR imaging approach has been introduced which can cope with \emph{non-rigid} motion. The motion-awareness is achieved by interleaved and repetitive measurements of the k-space center region while acquiring the remaining part of the k-space. This monitoring of the k-space center allows the observation of the motion which happens during the acquisition. The peripheral part of the k-space can thus be corrected for this observed motion. Due to the required re-acquisitions of the k-space center, the scanning time of over \SI{11}{\minute} and the subsequent reconstruction time of \SI{2}{\hour} exceed clinical applicability though.

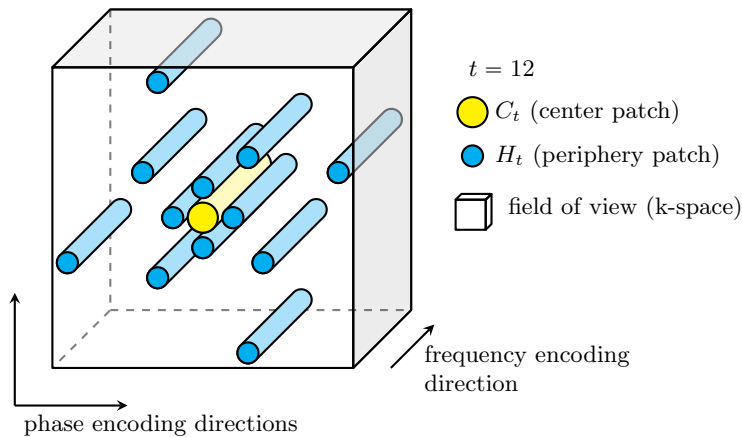
\begin{figure}[t]
  \begin{center}
  \input{sampling.tikz}
  \caption{Pseudo-random pattern in the phase encoding directions where the k-space center $C_t$ and a peripheral patch $H_t$ are sampled. Note that in the frequency encoding direction always the full k-space is sampled.}
  \label{fig:sampling}
\end{center}
\end{figure}

In this paper, we propose an accelerated motion-aware MR imaging approach to reduce the scanning and reconstruction time. The acquisition is divided into an initial phase where the motion is observed by acquiring the k-space center only. By generalizing from motion seen in this initial phase, the non-rigid motion to correct the remaining peripheral part of the k-space is inferred using a reduced version of the k-space center region. Thereby, one can almost half the scanning time from \SI{11}{\minute} to \SI{6}{\minute} and reduce the reconstruction time from \SI{2}{\hour} to \SI{3}{\minute}. Moreover, we extend the motion-aware approach~\cite{jud2018motion} by taking into account the systematic temporal shift which occurs in the patch alignment. Applying this shift correction reduces motion-induced blurring artifacts and results in a sharper image reconstruction.
In the experiments, we show that the average motion prediction error stays below \SI{2}{\milli\meter}. We further qualitatively show that the reconstruction quality is similar or higher than for the approach of~\cite{jud2018motion}.

There is a vast amount of literature concerning motion in MR imaging. For a comprehensive review about motion correction approaches in general and with a focus on prospective methods, we refer to~\cite{zaitsev2015motion} and ~\cite{maclaren2013prospective} respectively. Prospective methods react on motion during the image acquisition by correcting the pulse sequence accordingly. Generally, they required specialized equipment such as optical tracking systems and are thus difficult to apply in the clinical routine. Methods which rely on image or k-space navigators~\cite{white2010promo} require additional time in the sequence in order to acquire enough motion information for the correction. Because prospective methods have real-time constraints the computational budget is substantially restricted. 

In contrast to prospective approaches, retrospective approaches try to invert the motion effected changes in the acquired data after the acquisition has ended. Most retrospective and prospective methods are limited to rigid body motion. Methods which consider non-rigid motion in the reconstruction are usually based on gating~\cite{schmidt2011nonrigid} or binning of motion states~\cite{luo2017nonrigid,feng20185d}, limited to 2D~\cite{batchelor2005matrix,duffy2018retrospective}, or perform a piece-wise rigid approximation of the estimated motion~\cite{luo2017nonrigid}. Learning-based approaches try to reduce motion artifacts by directly converting a motion degenerated image into an artifact-free image~\cite{duffy2018retrospective,haskell2019network}. However, such approaches are limited to 2D~\cite{haskell2019network} or to isometric transformations~\cite{duffy2018retrospective}. Moreover, no time-resolved reconstructions are provided.

Current 4D approaches acquire partial image data over several quasi-periodic motion cycles and sort them retrospectively into time-resolved volumes using navigators which is called stacking~\cite{giger2018ultrasound,von20074d}. The validity of the resulting volumes is, however, difficult to assess. A review on such approaches can be found in~\cite{stemkens2018nuts}. What is unique in the herein proposed motion-aware methods among current 4D MR approaches is, that it provides continuous motion information while k-space is measured in 3D. Thus, binning, sorting and stacking is avoided. Instead of averaged motion cycles, the resulting time-resolved volumes --~a 4D MR image~-- contains the full variations of motion.

\section{Background}
In the motion-aware MR imaging method~\cite{jud2018motion} a consistent motion-corrected $\text{k-space}$ is reconstructed based on a specific sampling pattern where the k-space is divided into small patches (see \autoref{fig:sampling}). For each time point $t$, a part of the k-space $P_t\subset\mathbb{C}^3$ is sampled consisting of a patch within the center $C_t$ and a peripheral high-frequency patch $H_t$.

Before $P_t$ is accumulated into a consistent k-space, it is spatially aligned. Let $\mathcal{X}\subset\R^3$ denote the image domain. The $\text{\emph{non-rigid}}$ spatial motion $u_t:\mathcal{X}\rightarrow\R^3$ between a reference $P_r$ and any other $P_t$ is derived via image registration of the respective center patches.

The image $\bar{I}_r$ at the reference time point is finally reconstructed as follows:
\begin{align}
\label{eq:nonrigid_reconstruction}
 \bar{I}_r &= \mathcal{F}^{-1}\left(\bar{K}\right),\quad \bar{K} =W\sum_{t=1}^T W^t\mathcal{F}\Big(\mathcal{F}^{-1}\big(P_t\big)\circ u_t\Big),\\
 W(k_x) &=
 \begin{cases}
      \frac{1}{w_{k_x}} & w_{k_x}>0\\
      1 & \text{otherwise}
 \end{cases},~~
 w_{k_x}=\sum_{t=1}^T W^t(k_x), \quad W^t(k_x) =
 \begin{cases}
  1 & \vert P_t(k_x)\vert > 0\\
  0 & \text{otherwise},
 \end{cases}\nonumber
\end{align}
where $\bar{K}$ is the reconstructed k-space, $\mathcal{F}$ the Fourier transform, $W$ normalizes for overlapping patches and $W^t$ masks the patch after the non-rigid correction.

\section{Method}
\begin{figure}[t]
  \begin{center}
  \input{sequence.tikz}
  \caption{Illustration of the different sampling patterns for the \emph{standard} (upper row) and the \emph{accelerated} (lower row) motion-aware sequence. Center patches are visualized in yellow and peripheral patches in blue.}
  \label{fig:sequence}
\end{center}
\end{figure}
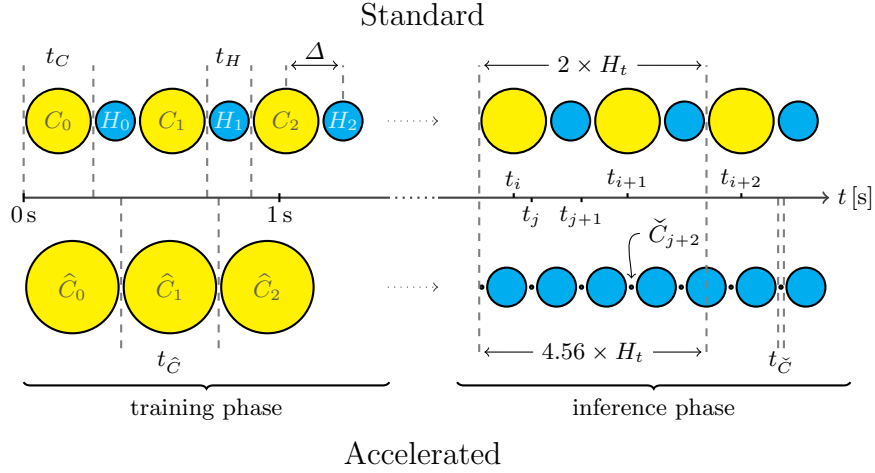

We present an accelerated version of the motion-aware approach~\cite{jud2018motion}, where the subject motion is modeled from an initial acquisition phase. The resulting model is used to infer the motion based on small-sized k-space centers in order to correct for motion in the peripheral patches. In \autoref{fig:sequence}, the \emph{standard}, and the proposed \emph{accelerated} motion-aware MR sampling strategies are visually compared.

\subsection{Motion Prediction using Cubic Regression}
In the alternating patch-wise sampling scheme the center of the k-space is measured within each time point. Although this ensures that the motion patterns can be recovered the patches are highly redundant. Our idea is to exploit this redundancy by predicting the motion fields having sampled only a tiny portion of the k-space center. After an initial phase, where the larger sized center is sampled only, the correlation between a tiny subset of this center and the recovered motion pattern is learned. In the subsequent \emph{inference} phase, these tiny centers are considered in the patch-wise sampling based on which the motion fields are predicted. The final overall acquisition time can thus be drastically reduced.

Let ${\widehat{C}}_t$ be the k-space centers in the initial \emph{training} phase which are used to recover $u_t$ using image registration. Let further $\widecheck{C}_t$ denote the tiny subset of the center and ${x_t\in\R^{d_\text{in}}}$ (input) the vectorized version of it with $d_\text{in}$ dimensions. Let us define ${y_t\in\R^{d_\text{out}}}$ (output) as the vectorized version of $u_t$ respectively with $d_\text{out}$ dimensions. Furthermore, we are given a training set $\left\{x_t,y_t\right\}_{t=1}^n$ of $n$ time points -- the training phase.
We define the cubic model
\begin{equation}
  y_t = \Psi z_t + \epsilon,\quad z_t=\{s_t, s_t^2, s_t^3\}
  \label{eq:model}
\end{equation}
where $s_t\in\R^{d_\text{pca}}$ are the scores of $x_t$ derived by projecting $x_t$ into the main modes of variation of the training set using principal component analysis while keeping $d_\text{pca}$ dimensions. The model $\Psi\in\R^{d_\text{out}\times 3d_\text{pca}}$ is a weight matrix with additive Gaussian noise $\epsilon$. It is identified in a least-squares sense.
Finally, a motion field $u_t$ is predicted given the tiny center $\widecheck{C}_t$ and the model weights $\Psi$.

\subsection{Systematic Shift Correction using Quadratic Interpolation}
\label{sec:shift_correction}
The motion-aware concept is based on the assumption that the motion which happens while acquiring $P_t$ is negligible. Nevertheless, applying the motion field $u_t$ directly to the peripheral patches induces a systematic temporal shift because the motion fields are derived using the center patches exclusively (see $\Delta$ in \autoref{fig:sequence}). Therefore, we propose to correct for this systematic shift using quadratic interpolation

\begin{equation}
u_{t\pm\Delta} = u_t + \dot{u}_t\left(t \pm \Delta\right) + \frac{1}{2}\ddot{u}_t\left(t \pm \Delta\right)^2,
\end{equation}
where the time derivatives $\dot{u}_t$ and $\ddot{u}_t$ are derived with central differences approximation. As the size of the patches is constant \emph{within} the different acquisition phases, the temporal center of $H_t$ is at $t+0.5$. Hence, we shift $u_t$ by $\Delta=0.5$.

\section{Experiments and Results}
\begin{table}[t]
 \centering
 \caption{The motion-aware sequence parameters used in the experiments.}
 \begin{tabular}[t]{lccr}
   Patch& Radius & Points & Time\\
   \hline
   $C_t$ & 6 & 109 & \SI{272.5}{\milli\second}\\
   $\widehat{C}_t$ & 10 & 305 & \SI{762.5}{\milli\second}\\
   $\widecheck{C}_t$ & 2 & 9 & \SI{22.5}{\milli\second}\\
   $H_t$ & 5 & 69 & \SI{172.5}{\milli\second}\\
   \multicolumn{3}{l}{Further parameters} & \\
  \hline
   \multicolumn{3}{l}{Repetition time:} & \SI{2.5}{\milli\second}\\
   \multicolumn{3}{l}{Echo time:} & \SI{1.0}{\milli\second}\\
   \multicolumn{3}{l}{Flip angle:} & \SI{5}{\degree}
 \end{tabular}
 \quad
 \begin{tabular}[t]{lr}
 Further parameters & \\
  \hline
  Bandwidth & \SI[quotient-mode=fraction, fraction-function=\dfrac]{1560}{\hertz/\pixel}\\

  Field of view & $400\times 400\times \SI{275}{\cubic\milli\meter}$\\
  Matrix & $128\times 128\times \SI{88}{\pixel}$\\
  Time points $T$ & 1500 \\
  Acquisition time (\emph{standard}) & \SI{11.1}{\minute}\\
  Acquisition time (\emph{training}) & \SI{76}{\second}\\
  Acquisition time (\emph{inference}) & \SI{4.5}{\minute}\\
  Acquisition time (\emph{accelerated}) & \SI{5.8}{\minute}
 \end{tabular}
 \label{tab:sequence}
\end{table}

We scanned the thorax and abdomen of $12$ volunteers under free breathing. We run the standard sequence~\cite{jud2018motion} on three different sites using Siemens MAGNETOM Prisma $3$ Tesla for six volunteers and Siemens MAGNETOM Aera $1.5$ Tesla for the other six volunteers. Detailed sequence parameters can be found in \autoref{tab:sequence}. For six of the volunteers, the accelerated motion-aware sequence has been run in addition. We distinguish between the following k-space accumulation strategies: static (no motion considered), non-rigid (standard non-rigid motion compensation~\cite{jud2018motion}), shift-corrected (non-rigid with shift correction) and accelerated (shift-corrected with variable center patch sizes $\widehat{C}_t$ and $\widecheck{C}_t$). For the accelerated reconstruction, the training phase lasts 100 time points and the inference phase 1400 time points. The different coil signals are combined in the spatial domain using root mean squares. The non-rigid image registration has been performed using AIRLab~\cite{sandkuhler2018airlab} applying a B-spline transformation model, an isotropic total variation regularizer on the motion field and the mutual information image-to-image metric. We further constrained the motion fields to be diffeomorphic and masked the overall minimization objective with a semi-automatically derived sliding-organ mask.

\subsubsection{Motion Prediction}
\begin{figure}[b]
\centering
  \input{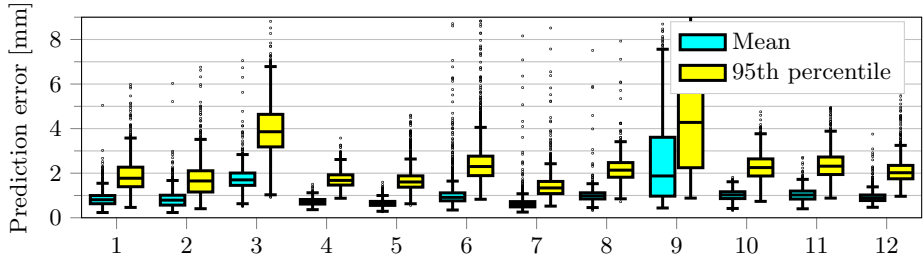}
  \caption{Boxplots with lower and upper quartile of the prediction errors for the 12 different volunteers: average error (cyan), 95th percentile (yellow). The whiskers extend to $1.5$ times of the interquartile range (IQR).}
  \label{fig:boxplots}
\end{figure}

In this experiment, we show the feasibility of the accelerated motion-aware method based on the standard acquisitions. The shift-corrected reconstruction serves as ground truth. We simulate the accelerated sampling strategy by considering full center patches $\widehat{C}_t:=C_t$ exclusively in a training phase and extracted the tiny patches $\widecheck{C}_t\subset C_t$ for the inference phase. Since the acquisition for the standard sequence extends over several minutes, we split the training phase into the leading and last time points to account for organ drift. We considered the leading ten principal components to build the motion model. 

In \autoref{fig:boxplots}, we analyze the motion prediction error of the model with respect to the ground truth. We distinguish between the average magnitude in displacement error (cyan) and the $95$th percentile (yellow) respectively. The average error stays below \SI{2}{\milli\meter} while $95\%$ of the prediction errors are on average smaller than \SI{4}{\milli\meter} with few exceptions. We observed a radical change in amplitude of the breathing pattern during the acquisition of Volunteer~9 which could explain the inferior prediction performance. Overall we can conclude that the generalization to motion patterns which have not been observed in the training phase might drop.

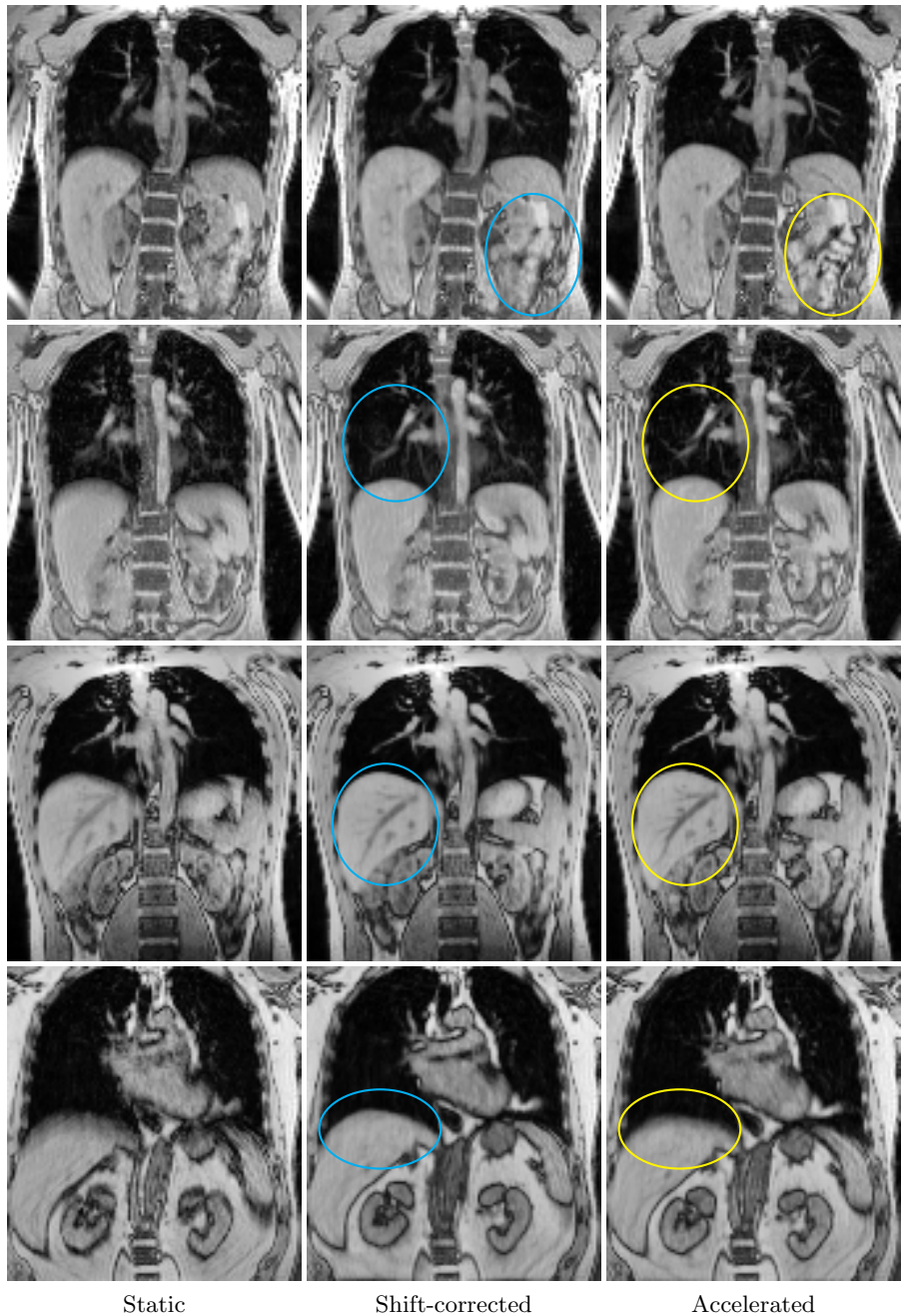
\begin{figure}[p]
\centering
\raisebox{0mm}{
\begin{tikzpicture}[node distance=2pt]
\node[inner sep=0pt] (first_left)  {\includegraphics[width=0.32\textwidth]{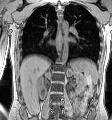}};
\node[inner sep=0pt, right= of first_left] (first_right) {\includegraphics[width=0.32\textwidth]{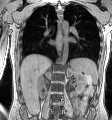}};
\node[inner sep=0pt, right= of first_right] (first_right2) {\includegraphics[width=0.32\textwidth]{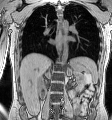}};
\node[inner sep=0pt, below= of first_left] (second_left) {\includegraphics[width=0.32\textwidth]{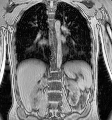}};
\node[inner sep=0pt, right= of second_left] (second_right) {\includegraphics[width=0.32\textwidth]{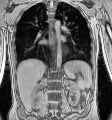}};
\node[inner sep=0pt, right= of second_right] (second_right2) {\includegraphics[width=0.32\textwidth]{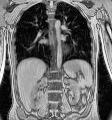}};
\node[inner sep=0pt, below= of second_left] (third_left) {\includegraphics[width=0.32\textwidth]{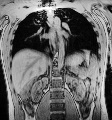}};
\node[inner sep=0pt, right= of third_left] (third_right) {\includegraphics[width=0.32\textwidth]{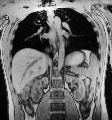}};
\node[inner sep=0pt, right= of third_right] (third_right2) {\includegraphics[width=0.32\textwidth]{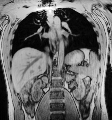}};
\node[inner sep=0pt, below= of third_left] (fourth_left) {\includegraphics[width=0.32\textwidth]{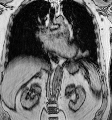}};
\node[inner sep=0pt, right= of fourth_left] (fourth_right) {\includegraphics[width=0.32\textwidth]{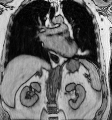}};
\node[inner sep=0pt, right= of fourth_right] (fourth_right2) {\includegraphics[width=0.32\textwidth]{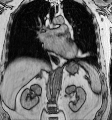}};

\node[below= of fourth_left]{Static};
\node[below= of fourth_right]{Shift-corrected};
\node[below= of fourth_right2]{Accelerated};

\def\xradiusa{18pt}
\def\yradiusa{23pt}
\def\xshifta{30pt}
\def\yshifta{-35pt}
\draw[thick, yellow] ([xshift=\xshifta, yshift=\yshifta]first_right2) circle [x radius=\xradiusa, y radius=\yradiusa];
\draw[thick, cyan] ([xshift=\xshifta, yshift=\yshifta]first_right) circle [x radius=\xradiusa, y radius=\yradiusa];

\def\xradiusb{20pt}
\def\yradiusb{22pt}
\def\xshiftb{-22pt}
\def\yshiftb{15pt}
\draw[thick, yellow] ([xshift=\xshiftb, yshift=\yshiftb]second_right2) circle [x radius=\xradiusb, y radius=\yradiusb];
\draw[thick, cyan] ([xshift=\xshiftb, yshift=\yshiftb]second_right) circle [x radius=\xradiusb, y radius=\yradiusb];

\def\xradiusc{20pt}
\def\yradiusc{23pt}
\def\xshiftc{-26pt}
\def\yshiftc{-8pt}
\draw[thick, yellow] ([xshift=\xshiftc, yshift=\yshiftc]third_right2) circle [x radius=\xradiusc, y radius=\yradiusc];
\draw[thick, cyan] ([xshift=\xshiftc, yshift=\yshiftc]third_right) circle [x radius=\xradiusc, y radius=\yradiusc];

\def\xradiusd{23pt}
\def\yradiusd{15pt}
\def\xshiftd{-28pt}
\def\yshiftd{-2pt}
\draw[thick, yellow] ([xshift=\xshiftd, yshift=\yshiftd]fourth_right2) circle [x radius=\xradiusd, y radius=\yradiusd];
\draw[thick, cyan] ([xshift=\xshiftd, yshift=\yshiftd]fourth_right) circle [x radius=\xradiusd, y radius=\yradiusd];

\end{tikzpicture}
}
\caption{Examples of coronal slices through reconstructed volumes. Rows: Volunteer $5,7,8$ and $12$. Columns: static and shift-corrected method using the standard acquisition. Last column: proposed reconstruction method using the accelerated acquisition. The colored ellipses mark regions of considerable differences.}
\label{fig:accelerated_reconstruction}
\end{figure}

\begin{figure}[h]
\centering
\raisebox{0mm}{
\begin{tikzpicture}[node distance=2pt]
\node[inner sep=0pt] (first_left)  {\includegraphics[width=0.32\textwidth]
{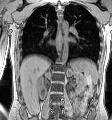}};
\node[inner sep=0pt, right= of first_left] (first_right) {\includegraphics[width=0.32\textwidth]{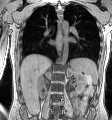}};
\node[inner sep=0pt, right= of first_right] (first_right2) {\includegraphics[width=0.32\textwidth]{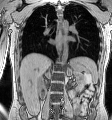}};
\node[inner sep=0pt, below= of first_left] (second_left) {\includegraphics[width=0.32\textwidth]{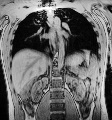}};
\node[inner sep=0pt, right= of second_left] (second_right) {\includegraphics[width=0.32\textwidth]{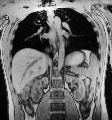}};
\node[inner sep=0pt, right= of second_right] (second_right2) {\includegraphics[width=0.32\textwidth]{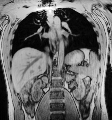}};

\node[below= of second_left]{Static};
\node[below= of second_right]{Shift-corrected};
\node[below= of second_right2]{Accelerated};

\end{tikzpicture}
}
\caption{Additional example coronal slices through the reconstructed volumes of Volunteers $4$ and $6$.}
\label{fig:accelerated_reconstruction2}
\end{figure}

\subsubsection{Accelerated Motion-Aware MR Imaging}
The accelerated acquisition \emph{and} reconstruction has been performed on six volunteers. The acquisition time has been reduced from \SI{11.1}{\minute} to \SI{5.8}{\minute} when compared to the standard acquisition (see \autoref{tab:sequence}). The reconstruction time could be reduced from \SI{2}{\hour} to \SI{3}{\minute} on average because the amount of time points where a 3D image registration is required is substantially reduced. We qualitatively assess the reconstruction of the two acquisition variants in \autoref{fig:accelerated_reconstruction} and \autoref{fig:accelerated_reconstruction2} for different subjects. In most cases, the accelerated method is even superior compared to the standard approach as it can be seen in \autoref{fig:accelerated_reconstruction}. This might be caused by the shorter time interval $\Delta$ in the inference phase. In the last case of  \autoref{fig:accelerated_reconstruction} as well as in the last case of \autoref{fig:accelerated_reconstruction2}, we observe minor blurring artifacts at the diaphragm. 

Note, that we cannot make a direct comparison between the shift-corrected and accelerated reconstruction method (middle and right column) because they are applied to different acquisitions (\emph{standard} and \emph{accelerated} respectively). 
As an example, for the reconstructions of Volunteer~6 (bottom row of \autoref{fig:accelerated_reconstruction2}) there are out-of-plane differences between the two visualized sliced. Furthermore, we emphasize that even an equivalent reconstruction quality is a major gain, because the acquisition and reconstruction time for the accelerated approach were considerably shorter.

\subsubsection{Systematic Shift Correction}
Finally, we tested the systematic shift correction (see Section~\ref{sec:shift_correction}) on $12$ standard and $6$ accelerated acquisitions. 

To quantify whether the shift correction increases the quality of the reconstructed image, we utilized the average total variation (TV) as a quality measure which quantifies sharpness. Note that no reference-based quality measure can be applied in this case because no ground-truth is available.

We investigated whether the TV increases with respect to the non-rigid reconstruction, by applying a one-sample t-test where the null hypothesis is no change in TV. We found a statistically significant increase in TV ($p=0.002$) for a significance level of $1\%$ with a large effect size (Cohen's d) of $d=0.86$.

\section{Conclusion}
We have presented an \emph{accelerated} motion-aware MR imaging approach which can cope with non-rigid motion and which yields a time-resolved volumetric MR image. The acquisition time has been reduced by a factor of two and the reconstruction time by two orders of magnitude compared to the standard approach~\cite{jud2018motion}. The acceleration is achieved by motion prediction using a motion model which is learned in an initial acquisition phase. In the experiments, we have reached a sub-pixel accuracy with an average motion prediction error which stays below \SI{2}{\milli\meter}. Moreover, the qualitative assessment shows equivalent or superior reconstruction quality. Changes in amplitude of the motion after the training phase may compromise the motion prediction though. Specific maximum inhalation and maximum exhalation breathing of the patient during the initial learning phase might be a solution to this problem, which will be investigated in future work.

\paragraph{Acknowledgements} 
This work was supported by the Swiss National Science Foundation, SNSF (320030\_163330/1) and the NVIDIA Corporation (with the donation of a GPU).

\bibliographystyle{splncs03}
\bibliography{bibliography}

\end{document}

%% file: sampling.tikz
\begin{tikzpicture}[scale=1, node distance=2pt]

  \pgfmathsetmacro{\offsetxy}{0.5}
  \pgfmathsetmacro{\offsetz}{0.5}
  \pgfmathsetmacro{\lengthxy}{1.5}
  \pgfmathsetmacro{\lengthz}{1.5}

  \draw[->,>=stealth, thick] (-\cubex-\offsetxy,-\cubey-\offsetxy,0) node [below right] {phase encoding directions} -- (-\cubex-\offsetxy+\lengthxy,-\cubey-\offsetxy,0); 
  \draw[->,>=stealth, thick] (-\cubex-\offsetxy,-\cubey-\offsetxy,0) -- (-\cubex-\offsetxy,-\cubey-\offsetxy+\lengthxy,0); 
  \draw[->,>=stealth, thick] (\offsetz,-\cubey,0) node [anchor=center, right=1em, align=left] {frequency encoding\\direction} -- (\offsetz,-\cubey,-\lengthz); 

  \def\cmtopt{28.45274}
  \begin{scope}[shift={(0,0,0)},
                    z={(\xx*\cmtopt pt,\xy*\cmtopt pt)},
                    x={(\yx*\cmtopt pt,\yy*\cmtopt pt)},
                    y={(\zx*\cmtopt pt,\zy*\cmtopt pt)}]

   \foreach \sx/\sy in {0.6/1.8,
                        0.6/-0.8,
                        0.4/0,
                        0/-0.4,
                        0.8/0.6,
                        -1.8/0.6
                        }{
    \tdcylxy{\cubex*0.5+\sx}{-\cubey*0.5+\sy}{0}{\rb}{\cubez*0.5}{side1}{base1} 
   }
    \tdcylxy{\cubex*0.5}{-\cubey*0.5}{0}{\ra}{\cubez*0.5}{side}{base} 

  \end{scope}

    \draw[thick, line join=round] (0,0,0) -- ++(-\cubex,0,0) -- ++(0,-\cubey,0) -- ++(\cubex,0,0) -- cycle;
    \draw[thick, fill=faceb,opacity=0.5, line join=round] (0,0,0) -- ++(0,0,-\cubez) -- ++(0,-\cubey,0) -- ++(0,0,\cubez) -- cycle;
    \draw[thick, line join=round] (0,0,0) -- ++(0,0,-\cubez) -- ++(0,-\cubey,0) -- ++(0,0,\cubez) -- cycle;
    \draw[thick, fill=facec,opacity=0.5, line join=round] (0,0,0) -- ++(-\cubex,0,0) -- ++(0,0,-\cubez) -- ++(\cubex,0,0) -- cycle;
    \draw[thick, line join=round] (0,0,0) -- ++(-\cubex,0,0) -- ++(0,0,-\cubez) -- ++(\cubex,0,0) -- cycle;
    \draw[thick, dashed, opacity=0.5, line join=round] (-\cubex,0,-\cubez) -- (-\cubex,-\cubey,-\cubez) -- (-\cubex,-\cubey,0);
    \draw[thick, dashed, opacity=0.5, line join=round] (-\cubex,-\cubey,-\cubez) -- (0,-\cubey,-\cubez);

    \node (time) at (\cubex*0.3,-\cubey*0.1,-\cubez*0.5){};
    \node[right= of time, xshift=-10pt] (ttext) {$t=12$};
    \node[below=10pt of time] (center){};
    \filldraw[thick, black, fill=base] (center) circle[radius=\ra];
    \node[right= of center] (ctext) {$C_t$ (center patch)};

    \node[below=10pt of center] (periphery){};
    \filldraw[thick, black, fill=base1] (periphery) circle[radius=\rb];
    \node[right= of periphery] (ptext) {$H_t$ (periphery patch)};
    
    \node[below=10pt of periphery, xshift=5pt] (cuboid){};
    \begin{scope}[shift={(cuboid)}, scale=0.1]
      \draw[thick, line join=round] (0,0,0) -- ++(-\cubex,0,0) -- ++(0,-\cubey,0) -- ++(\cubex,0,0) -- cycle;
      \draw[thick, line join=round] (0,0,0) -- ++(0,0,-\cubez) -- ++(0,-\cubey,0) -- ++(0,0,\cubez) -- cycle;
      \draw[thick, line join=round] (0,0,0) -- ++(-\cubex,0,0) -- ++(0,0,-\cubez) -- ++(\cubex,0,0) -- cycle;
    \end{scope}
    \node[right= of cuboid, yshift=-3pt] (cubtext) {field of view (k-space)};

  \begin{scope}[shift={(0,0,0)},
                    z={(\xx*\cmtopt pt,\xy*\cmtopt pt)},
                    x={(\yx*\cmtopt pt,\yy*\cmtopt pt)},
                    y={(\zx*\cmtopt pt,\zy*\cmtopt pt)}]
    \foreach \sx/\sy in {-0.4/0,
                        0/0.4,
                        -0.6/0.8,
                        -0.8/-0.6,
                        1.8/-0.6,
                        -0.6/-1.8
                        }{
     \tdcylxy{\cubex*0.5+\sx}{-\cubey*0.5+\sy}{0}{\rb}{\cubez*0.5}{side1}{base1} 
    }

  \end{scope}

\end{tikzpicture}

%% file: sequence.tikz
\begin{tikzpicture}[scale=3.4]
\def\offsetn{0.3} 
\def\offsetl{0.35} 
\def\tcn{0.2725} 
\def\tcs{0.0225} 
\def\tcl{0.38125} 
\def\tpn{0.1725} 
\def\dotr{0.002} 
\def\smalldelta{0.01}

\foreach \i in {0,1,2,4,5,6} {
  \def\xc{\i*\tcn + \i*\tpn + \tcn/2}
  \draw[thick, fill=yellow] (\xc,\offsetn) circle [radius=\tcn/2-\smalldelta];
  
  \ifthenelse{\i<3}
    {\node[black!75!yellow] at (\xc,\offsetn) {$C_{\i}$};}
    {}
  ;


  \def\xp{\i*\tcn + \i*\tpn + \tcn + \tpn/2}
  \draw[thick, fill=cyan] (\xp,\offsetn) circle [radius=\tpn/2-\smalldelta];
  
  \ifthenelse{\i<3}
    {\node[white!85!cyan] at (\xp,\offsetn) {$H_{\i}$};}
    {}
  ;
  
}

\draw[thick, gray, dashed] (0,0) -- (0,\offsetn+\tcn/1.25);
\draw[thick, gray, dashed] (\tcn,0) -- (\tcn,\offsetn+\tcn/1.25);
\node[above] at (\tcn/2,\offsetn+\tcn/1.5) {$t_C$};

\draw[thick, gray, dashed] (2*\tcn + \tpn,0) -- (2*\tcn + \tpn,\offsetn+\tcn/1.25);
\draw[thick, gray, dashed] (2*\tcn + 2*\tpn,0) -- (2*\tcn + 2*\tpn,\offsetn+\tcn/1.25);
\node[above] at (2*\tcn + 1.5*\tpn,\offsetn+\tcn/1.5) {$t_H$};

\def\offsetd{2.5*\tcn + 2*\tpn}
\def\shifttwo{\tcn/4+\tpn/4}
\node (b) at (\offsetd+\shifttwo, \offsetn+\tcn) {$\Delta$};
\draw[<->,above] (\offsetd+\smalldelta*2, \offsetn+\tcn/1.25) -- (\offsetd+0.5*\tcn + 0.5*\tpn-\smalldelta*2, \offsetn+\tcn/1.25);

\draw[thick, gray, dashed] (2.5*\tcn + 2*\tpn,\offsetn+\tcn/2) -- (2.5*\tcn + 2*\tpn,\offsetn+\tcn/1.25);
\draw[thick, gray, dashed] (3*\tcn + 2.5*\tpn,\offsetn+\tpn/2) -- (3*\tcn + 2.5*\tpn,\offsetn+\tcn/1.25);

\draw[thick, gray, dashed] (\tcl,0) -- (\tcl,-\offsetl-\tcl/1.5);
\draw[thick, gray, dashed] (2*\tcl,0) -- (2*\tcl,-\offsetl-\tcl/1.5);
\node[below] at (1.5*\tcl,-\offsetl-\tcl/1.75) {$t_{\widehat{C}}$};

\foreach \i in {0,...,2} {
  \def\xc{\i*\tcl + \tcl/2}
  \draw[thick, fill=yellow] (\xc,-\offsetl) circle [radius=\tcl/2-\smalldelta];
  \node[black!75!yellow] at (\xc,-\offsetl) {$\widehat{C}_{\i}$};
}

\def\offsets{3*\tcn + 4*\tpn + \tcn}
\foreach \i in {0,...,6} {
  \def\xc{\offsets + \i*\tcs + \i*\tpn + \tcs/2}
  \draw[thick, fill=yellow] (\xc,-\offsetl) circle [radius=\tcs/2-\smalldelta/2];

  \def\xp{\offsets + \i*\tcs + \i*\tpn + \tcs + \tpn/2}
  \draw[thick, fill=cyan] (\xp,-\offsetl) circle [radius=\tpn/2-\smalldelta];
}
\def\offsetsl{\offsets + \tcs/2 - \smalldelta} 
\draw[dashed, thick, gray] (\offsetsl, -\offsetl-\tcl/1.5) -- (\offsetsl,\offsetn+\tcn/1.25);
\def\offsetsll{\offsets + 2*\tcn + 2*\tpn} 
\draw[dashed, thick, gray] (\offsetsll, -\offsetl-\tcl/1.5) -- (\offsetsll,\offsetn+\tcn/1.25);

\def\clabel{\offsets + 4*\tcs + 3.85*\tpn + \tcs/2}
\def\xc{\offsets + \i*\tcs + \i*\tpn + \tcs/2}
\node (z1) at (\clabel,-\offsetl/2.5){$\widecheck{C}_{j+2}$};
\def\cclabel{\offsets + 3*\tcs + 3*\tpn + \tcs/2}
\node (z2) at (\cclabel, -\offsetl){};
\draw[->] (z1) to[out=180,in=90] (z2);

\node (a) at (\offsetsl + \tcn + \tpn, -\offsetl-\tcl/1.5) {$4.56 \times H_t$};
\draw[<-,below] (\offsetsl+\smalldelta*2, -\offsetl-\tcl/1.5) -- (a);
\draw[->,below] (a) -- (\offsetsll-\smalldelta*2, -\offsetl-\tcl/1.5);

\node (b) at (\offsetsl + \tcn + \tpn, \offsetn+\tcn/1.25) {$2 \times H_t$};
\draw[<-,above] (\offsetsl+\smalldelta*2, \offsetn+\tcn/1.25) -- (b);
\draw[->,above] (b) -- (\offsetsll-\smalldelta*2, \offsetn+\tcn/1.25);

\def\offsetslll{\offsets + 6*\tcs + 6*\tpn}
\draw[dashed, thick, gray] (\offsetslll, 0) -- (\offsetslll,-\offsetl-\tcl/1.5);
\def\offsetsllll{\offsets + 7*\tcs + 6*\tpn}
\draw[dashed, thick, gray] (\offsetsllll, 0) -- (\offsetsllll,-\offsetl-\tcl/1.5);
\def\offsetslllll{\offsets + 6.5*\tcs + 6*\tpn}
\node[below] at (\offsetslllll,-\offsetl-\tcl/1.75) {$t_{\widecheck{C}}$};

\draw [thick, darkgray] (0,0) -- (3.2*\tcn + 3.2*\tpn,0){};
\draw [thick, dotted, darkgray] (3.2*\tcn + 3.2*\tpn,0) -- (4.25*\tcl,0){};
\draw [->, thick, darkgray] (4.25*\tcl,0) -- (3.15,0){};

\foreach \x in {0,1}{
  \node [below] at (\x,0) {\SI{\x}{\second}};
  \draw[thick] (\x,-0.5pt) -- (\x,0.5pt);
}

\foreach \x/\y/\t in {4.5/4/$t_i$, 5.5/5/$t_{i+1}$, 6.5/6/$t_{i+2}$}{
  \node [above] at (\x*\tcn+\y*\tpn,0) {\t};
  \draw[thick] (\x*\tcn+\y*\tpn,0) -- (\x*\tcn+\y*\tpn,0.5pt);
}

\foreach \x/\t in {1/$t_j$, 2/$t_{j+1}$}{
  \node [below] at (\offsetsl+\smalldelta+\x*\tcs+\x*\tpn,0) {\t};
  \draw[thick] (\offsetsl+\smalldelta+\x*\tcs+\x*\tpn,0) -- (\offsetsl+\smalldelta+\x*\tcs+\x*\tpn,-0.5pt);
}

\node[right] at (3.15,0){$t$\,[s]};

\draw [->, dotted, darkgray] (3.2*\tcn + 3.2*\tpn, \offsetn) -- (4.25*\tcl, \offsetn);
\draw [->, dotted, darkgray] (3.2*\tcn + 3.2*\tpn, -\offsetl) -- (4.25*\tcl, -\offsetl);

\draw [thick, 
    decoration={ brace, mirror, raise=0.4cm},
    decorate] 
    (0,-\offsetl-\tcl/1.5) -- (3.75*\tcl,-\offsetl-\tcl/1.5) 
    node [pos=0.5,anchor=north,yshift=-0.5cm] {training phase}; 

\draw [thick, 
    decoration={ brace, mirror, raise=0.4cm},
    decorate] 
    (\offsetsl-0.5*\tpn,-\offsetl-\tcl/1.5) -- (\offsetsl+7.5*\tpn+7*\tcs,-\offsetl-\tcl/1.5) 
    node [pos=0.5,anchor=north,yshift=-0.5cm] {inference phase};

\node[below, black] at (3.7*\tcn + 3.2*\tpn,-\offsetl-\tcl*1.5) {\large Accelerated};
\node[above, black] at (3.7*\tcn + 3.2*\tpn,\offsetn+\tcn*1.25) {\large Standard};

\end{tikzpicture}